\documentclass[manuscript,screen]{acmart}
\usepackage{subcaption}

\usepackage{xcolor}

\usepackage{booktabs}
\usepackage{multirow}

\newif\ifshowedits
\showeditstrue      

\definecolor{zekunColor}{HTML}{000000}   
\definecolor{huiyongColor}{HTML}{000000}  
\newcommand{\zekun}[1]{\ifshowedits\textcolor{zekunColor}{#1}\else#1\fi}

\newcommand{\edit}[1]{\ifshowedits\textcolor{black}{#1}\else#1\fi}

\AtBeginDocument{%
  }

\setcopyright{acmlicensed}
\copyrightyear{2018}
\acmYear{2018}
\acmDOI{XXXXXXX.XXXXXXX}
\acmConference[Conference acronym 'XX]{Make sure to enter the correct
  conference title from your rights confirmation email}{June 03--05,
  2018}{Woodstock, NY}
\acmISBN{978-1-4503-XXXX-X/2018/06}

\begin{document}
\raggedbottom
\title{Gaze-Informed Proactive AI Assistance for Children Exploring Picture Books}

\author{Zekun Wu}
\email{wuzekun@cs.uni-saarlaned.de}
\orcid{0000-0002-5233-2352}
\affiliation{%
  \institution{Saarland University, Saarland Informatics Campus}
  \streetaddress{Campus, 66123 Saarbrücken}
  \city{Saarbrücken}
  \state{Saarland}
  \country{Germany}
  \postcode{66123}
}

\author{Man Su}
\email{m.su@iwm-tuebingen.dee}
\affiliation{%
  \institution{Tübingen Digital Teaching Lab (TüDiLab) , Leibniz-Institut für Wissensmedien (IWM Tübingen)}
  \city{Tübingen}
  \state{Baden-Württembergs}
  \country{Germany}
}

\author{Huiyong Li}
\email{li.huiyong.194@m.kyushu-u.ac.jp}
\affiliation{%
  \institution{Research Institute for Information Technology, Kyushu University}
  \city{Fukuoka}
  \country{Japan}
}

\author{Tomohiro Nagashima}
\email{nagashima@cs.uni-saarland.de}
\affiliation{%
  \institution{Saarland University, Saarland Informatics Campus}
  \city{Saarbrücken}
  \country{Germany}}

\author{Anna Maria Feit}
\email{feit@cs.uni-saarland.de}
\orcid{0000-0003-4168-6099}
\affiliation{%
  \institution{Saarland University, Saarland Informatics Campus}
  \streetaddress{E1 7, 66123 Saarbrücken}
  \city{Saarbrücken}
  \country{Germany}}


\renewcommand{\shortauthors}{Wu et al.}

\begin{abstract}
Proactive assistance with large language models has received growing attention in the HCI community. We explore how proactive assistance can support children during picture exploration where children may not formulate explicit requests, and where assistance risks steering their exploration rather than following it. We develop a gaze-based storytelling assistant that decides when to assist from the stability of a child's attention, and what to narrate by extending from where the child is already looking. In a within-subject study with 22 children, gaze-informed assistance held children's attention on the narrated area longer and led them to the related area more often than an LLM-baseline. Children, parents, and a kindergarten teacher preferred it, finding it more responsive to how children explored the picture. Our work suggests that gaze offers a way to make proactive assistance follow rather than direct a user's activity, in settings where needs are difficult to anticipate or articulate.

\end{abstract}

\ccsdesc[500]{Do Not Use This Code~Generate the Correct Terms for Your Paper}
\ccsdesc[300]{Do Not Use This Code~Generate the Correct Terms for Your Paper}
\ccsdesc{Do Not Use This Code~Generate the Correct Terms for Your Paper}
\ccsdesc[100]{Do Not Use This Code~Generate the Correct Terms for Your Paper}

\keywords{Do, Not, Us, This, Code, Put, the, Correct, Terms, for,
  Your, Paper}

\received{20 February 2007}
\received[revised]{12 March 2009}
\received[accepted]{5 June 2009}

\maketitle

\begin{figure*}[t]
  \centering
  \includegraphics[width=\textwidth]{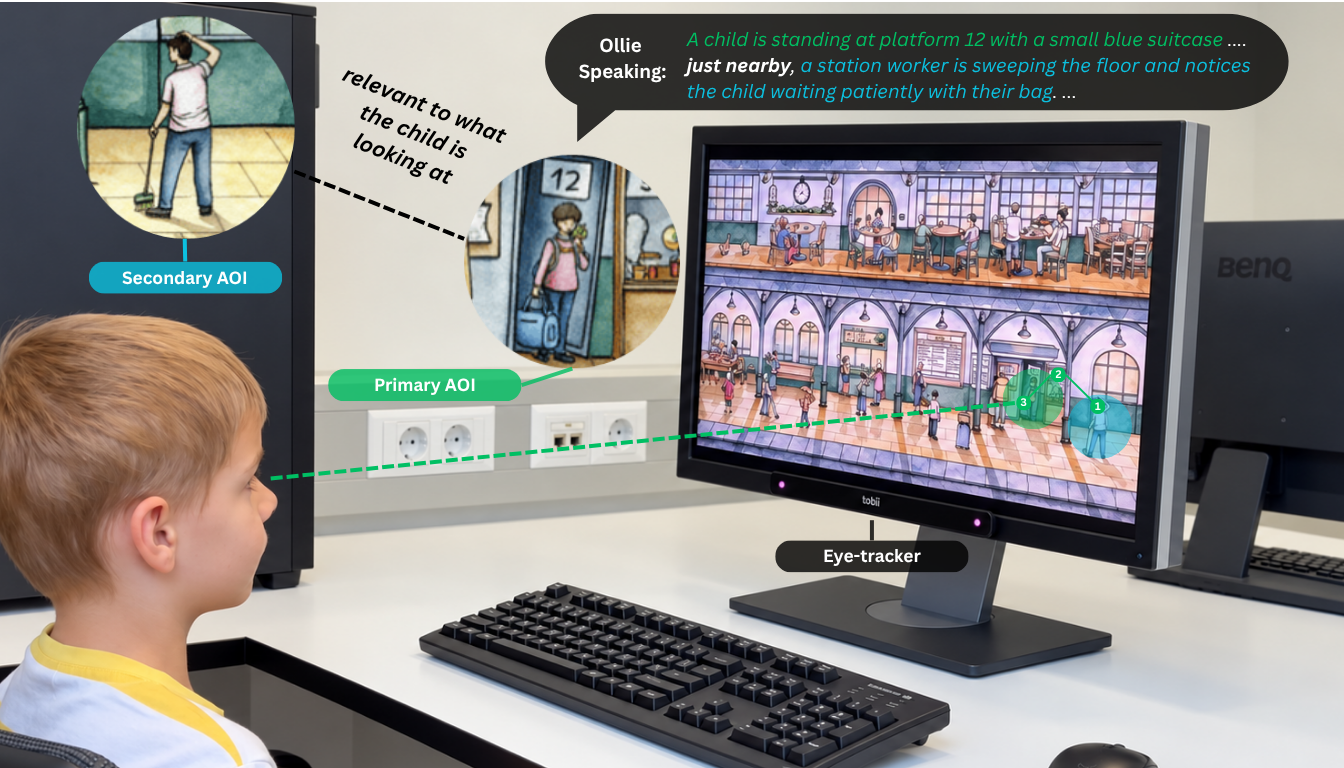}
  \caption{In this work, we explore how gaze information can be used to improve LLM-based proactive assistant, specifically in the context of open-ended activities. We develop Ollie to guide children with narrative descriptions during picture book exploration. Ollie starts from the child’s current focus (fixation 3 and green circle) and then introduces a contextually relevant area the child has previously fixated (fixation 1 and blue circle), producing a more coherent narrative link between the two regions that follows the child's attention. }
  \label{fig:teaser}
\end{figure*}

\section{Introduction}
\label{sec:intro}
The advance of large language models (LLMs) has created new opportunities to use AI as a \emph{proactive} partner that offers timely, context-aware assistance by anticipating users’ needs in a wide range of activities \cite{liu2025proactive,yang2025socialmind}. While reactive assistance requires users to stop their ongoing task and formulate explicit commands or queries, typically in the form of written or spoken prompts,
proactive assistance can provide support without requiring the user to interrupt their task to initiate it and can address needs that are difficult or cumbersome to articulate verbally \cite{zhao2025proactiveva,yang2025socialmind}. This is particularly relevant for a wider adoption of AI-based assistance as formulating effective prompts remains a known barrier for many users, who struggle to translate their goals into queries that elicit useful responses \cite{chen2025need,subramonyam2024}, given the functional flexibility of LLM-based assistants: users must not only articulate their need, but also identify which of the assistant's many possible capabilities would best support them at a given moment \cite{subramonyam2024, lehmann2024functionalflexibilitygenerativeai}.

\zekun{Among various user groups, proactive assistance can be especially useful for children, including preschool and early elementary-school children, who may not yet have sufficient skills to formulate clear prompts, request specific help, or even recognize what kind of support they might benefit from during an activity \cite{veenman2004relation}. However, proactive guidance in open-ended activities must be carefully aligned with children’s ongoing interests without narrowing children’s spontaneous exploration \cite{trawick2011good,voysey2026agency}.} Prior work has shown that children often communicate their interests and intentions through non-verbal behaviors, including gaze patterns \cite{liu2025eye} and facial expressions \cite{wu2020understanding}, which can indicate internal states such as attention, engagement, confusion, curiosity, and emotional response beyond verbal exchange. \zekun{Yet little is known about how such signals can be used to provide assistance that follows and extends children’s exploration without overriding their ongoing interests.}


One common open-ended task that children engage with in their daily lives is picturebook exploration. Picture books present various kinds of illustrations that carry a substantial share of the storytelling, often without written text. Picture books are popular among families, primary schools, and childcare facilities as they support  children’s visual literacy \cite{zacks2020event}, narrative understanding \cite{unal2019children}, and emotional interpretation \cite{pelz2020signature}. In practice, children often benefit from adult support during such activities, for example through storytelling, guiding questions, or open dialogue with peers \cite{maine2021using}. \edit{Nevertheless, such support is not always available or easy to provide \cite{dietz2024contextq}. Shared reading requires active communication and participation from both adults and children, while working parents may have limited time for these interactions \cite{lee2022can}. Moreover, even when adult support is available,} providing such support can be challenging for parents and teachers \cite{dietz2024contextq}, who may struggle to come up with engaging stories or questions that align with children’s moment-to-moment attention and interests \cite{zhang2022storybuddy}. This challenge has motivated a range of efforts to use AI to support children’s engagement during reading, including picturebook storytelling \cite{zhang2022storybuddy}, visual drawing \cite{zhang2022storydrawer,ye2024storypark}, and interactive narratives \cite{cheng2025oak}. \zekun{However, most existing approaches depend on explicit input from children or adults. It therefore remains unclear how an autonomous system can provide real-time narration that is grounded in children’s ongoing behavior and aligned with their self-directed exploration.}

\zekun{Motivated by this challenge, we investigate whether children’s eye gaze can support proactive narration that both reduces the need for explicit input and remains aligned with their self-directed picture exploration. Specifically, we use gaze as an approach to determine the suitable moments for intervention based on the child’s ongoing inferred (dis)engagement during picture exploration, and what story to tell about the picture so that it matches a child's interest and engages them in the exploration process.}

As illustrated in the top right part of \autoref{fig:teaser}, we developed Ollie, a 
LLM-based assistant that accompanies children in open-ended picturebook exploration. Rather than waiting for children to explicitly request help, Ollie uses gaze information to determine when to proactively initiate narration and what to talk about. When Ollie detects that a child is attending to a particular picture area, it uses that area as the starting point for narration. It then chooses a second area by considering how closely its content and location relate to the current area of attention, as well as how recently the child looked at it, and connects the two areas in the narration.

To evaluate the effectiveness of Ollie, we conducted a within-subject experiment with children, comparing gaze-based assistance with an LLM-baseline condition without gaze information. In the baseline condition, both the primary and secondary areas of interest  (AOIs) were randomly selected from the picture while using the same narration structure. We examined two aspects: how effectively the assistance guided children’s attention to the narrated AOIs, and how children and accompanying adults perceived the two approaches. Our results show that gaze-based assistance provided more effective visual guidance than the baseline condition. The gaze-based assistance was also preferred by most children, parents, and the participating teacher, who perceived it as better aligned with  \zekun{where children were looking and how they were exploring the picture} compared with the baseline condition.

Our work contributes the following:

\begin{itemize}
    \item \zekun{We develop and validate a method for using children’s visual attention as an implicit input for grounding LLM-generated narration during open-ended  exploration tasks.}
    \item We share empirical evidence that gaze-informed narration can better sustain children’s attention, guide them toward related picture regions, and make the assistance feel more \zekun{responsive to children’s ongoing visual exploration.}
\end{itemize}

Beyond picture exploration, our findings also suggest that eye gaze can be a promising signal for proactive AI assistance in open-ended exploration activities, where users' needs are difficult to anticipate or articulate.

\section{Background and Related Work}

\subsection{Digital Support for Children's Picture Book Reading and Exploration}


\edit{Picture book reading plays an important role in children’s language development and long-term cognitive growth \cite{bruner1991narrative,kress2005before,greenhoot2014more,rabiner2016predicting,xing2025effects}. It supports language acquisition \cite{greenlee1996interactive}, narrative comprehension \cite{xing2025effects}, memory \cite{greenhoot2014more}, creativity \cite{weeks2013power}, and overall school readiness \cite{wang2025picture}. To facilitate children’s picture reading, digital technologies have been explored as a way to extend traditional printed books, with reported benefits for maintaining attention, vocabulary learning, and reading comprehension \cite{alper2017visualization,rubegni2021girl}. Early efforts mainly added multimedia features to storybooks, such as sound effects, animations, colorful pictures, and varied text styles \cite{greenlee1996interactive,underwood1998children}. More recent systems provide support in multimodal, adaptive, and socially engaging ways \cite{liu2024potential,wang2019impact,csimcsek2021comparing,lin2022color,lyu2024emooly,vargas2025exploring}. These include multimedia narration and animation \cite{wang2019impact}, interactive elements and gamification for engagement \cite{son2024effects}, remote shared-reading tools for social-emotional connection \cite{lin2022color}, and personalized learning environments supported by adaptive systems \cite{sun2024storychat,michaelis2017someone,sun2022bilingual}.}


However, one persistent challenge remains for these support designs: children do not always clearly express what they are interested in, confused by, or ready to explore next \cite{nojavanasghari2016future,eckstein2017beyond,lodge2018understanding,ohnishi2024curiosity,kim2026choosing}. This has motivated research that uses behavioral signals, such as eye gaze, to analyze children’s attention and its effect on engagement or comprehension to inform the digital support design  \cite{eckstein2017beyond,skibbe2018preschoolers,liao2020electronic,liu2024interference}. 
For example, Skibbe et al. showed that children paid the most attention to text when it was both read aloud and visually highlighted \cite{skibbe2018preschoolers}. Eng et al. found that the design of picture books with removed extraneous details, leaving only illustrations relevant to the text, can significantly help improve child attention and reading comprehension \cite{eng2020keep}. Similarly, Liu et al. found that animations in digital picture books improved comprehension only when they were highly relevant to key story elements \cite{liu2024interference}. 

With the rise of LLMs, AI systems have become increasingly capable of supporting children’s recreational and learning activities \cite{zhang2022storybuddy,zhang2022storydrawer,bus2025interactive,cheng2025oak}. Still, without a careful understanding of children’s needs and thoughtful interaction design, these systems may provide support that is poorly timed, poorly grounded, \zekun{or overly directive. In open-ended activities, AI assistance that fails to consider children’s own interests and autonomy may constrain children's spontaneous exploration, limit their curiosity and engagement, and reduce their sense of control over how the activity unfolds \cite{bus2025interactive,kurian2025ai,berson2025innovating,zhu2025fostering,voysey2026agency}}. \zekun{Indeed, Berson et al. \cite{berson2025innovating}, warn that overly prescriptive AI may steer children along predetermined paths rather than supporting autonomous decision-making and self-directed learning. Cheng et al.\cite{cheng2025oak} emphasize the importance of balancing learner-directed activity with structure introduced by the AI system. Zhu et al. show that involving children in the design of AI systems can foster aspects of their autonomy and shift them from passive users toward more active engagement with AI\cite{zhu2025fostering}. }



These findings encourage us to further explore how implicit behavioral signals, such as eye gaze, can be used in the design of AI-based digital support for children. Prior work has addressed children's autonomy mainly at design time, by involving them in shaping the system. The design challenge we take up here arises at interaction time, where the system decides continuously and without explicit input from the child. Rather than asking what multimedia features should be added to a picture book, the question is how AI can provide timely and relevant support that follows the child's own interest rather than imposing a predetermined path. This motivates us to develop a gaze-informed AI assistance approach that adapts to children’s moment-to-moment visual attention while keeping their own exploration behavior as the primary driver of the interaction.

\subsection{LLM-Based Proactive Assistance and the Role of Eye Gaze}

The concept of proactive AI assistance has a long history \cite{horvitz1999principles}. One of the earliest approaches is the mixed-initiative interaction, where both the user and the system can take the initiative to guide a process or achieve a goal, particularly in productivity tool design \cite{horvitz1999principles,horvitz2013lumiere}. Recent advances in LLMs have made proactive AI assistance more feasible, as these models can interpret richer contextual information and provide support through natural language, but two key challenges remain: deciding \textit{when} to intervene and \textit{what} assistance to provide. Prior work has approached these challenges by reading cues from the ongoing user interaction: conversational context \cite{liu2025proactive}, idle states \cite{chen2025need}, and pauses or annotations \cite{zhao2025proactiveva} have been used to judge \textit{when} assistance may be appropriate, while task and interaction context \cite{yang2025socialmind,chen2025need,zhao2025proactiveva} have helped shape \textit{what} assistance to provide. Gaze is particularly promising in this regard, as it provides a continuous and implicit signal of how the user’s attention unfolds during an ongoing activity.

Only recently have researchers begun exploring how to combine gaze data with LLM-based assistance~\cite{danry2026gaze,bae2026llms,liu2026attentivelearn}. Danry et al. developed an LLM-driven gaze-aware assistant to support reading comprehension, but the assistance was provided reactively, i.e. only after the users completed each passage rather than during reading in real time \cite{danry2026gaze}. Bae et al. also focused on reading support, moving the granularity of gaze-based assistance to the sentence level through deviation-based detection over gaze features \cite{bae2026llms}. Liu et al. proposed an immersive learning system that generated quizzes based on students’ attention distribution across lecture sections, and showed that gaze-based personalized assistance could improve attention management, motivation, engagement, and comprehension \cite{liu2026attentivelearn}. While these studies demonstrate the potential of gaze for grounding and personalizing LLM-based assistance, less is known about how gaze can serve as an implicit signal for proactive assistance, informing both \textit{when} to intervene and \textit{what} assistance to provide during an ongoing activity. This is particularly relevant for children, whose moment-to-moment attention and potential interests may be reflected in their visual behavior rather than expressed through explicit requests.

\section{Ollie's Gaze-Informed Assistance Pipeline}
\label{sec:method}

This section presents the design goals of Ollie and describes how they are implemented through its gaze-informed proactive assistance pipeline.

\subsection{Design Goals and System Overview}
\label{sec:design-goals}
\zekun{Motivated by the literature discussed above, we formulate two design goals that guide the design of our gaze-based proactive assistance tool  for open-ended picture book exploration:}

\textbf{DG1: Ground assistance in children’s current attention.}
\zekun{Ollie should provide assistance only when a child's attention is stable on a specific AOI and ground the narration in that AOI. The timing should reflect the child’s attentional state and follow their exploration  rather than a fixed schedule or requiring an explicit request.}

\textbf{DG2: Follow the child's exploration rather than direct it.}
\zekun{Ollie should provide assistance that follows and extends the child’s ongoing gaze behavior, rather than steering attention towards regions the system considers important. The child should remain the primary driver of the exploration, with the narration responding to their potential interests instead of imposing a predetermined path.}

\begin{figure}[t]
\centering
\includegraphics[width=6.0in]{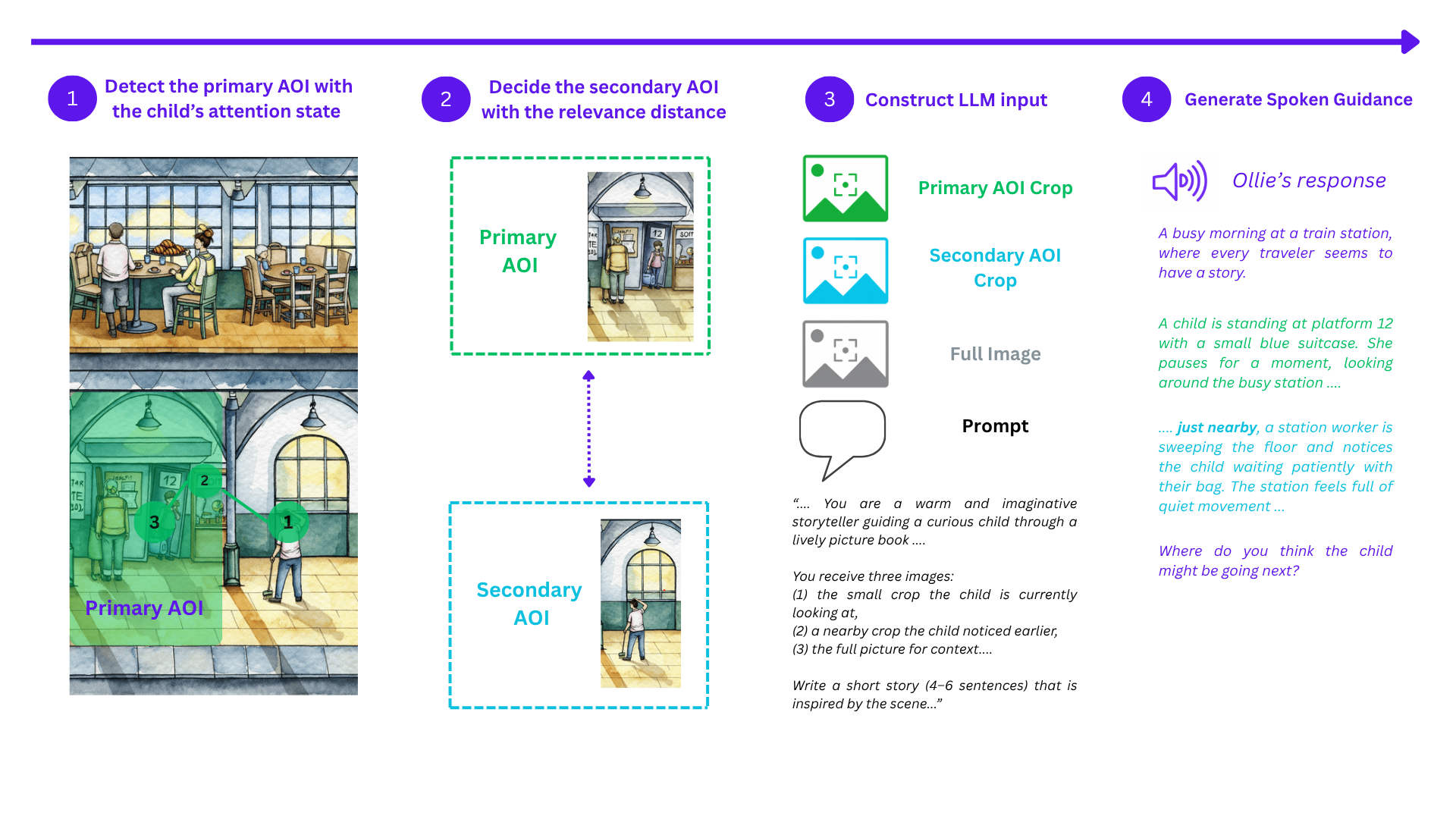}
\caption{Overview of the gaze-informed proactive assistance pipeline. Ollie uses gaze-based attention detection to identify the child’s current primary AOI, applies relevance-based AOI selection to choose a related but unassisted secondary AOI, and then prompts the LLM with the two AOIs and the full picture to produce spoken storytelling assistance.}
\Description{A four-step pipeline from a child's gaze to spoken storytelling assistance. First, gaze-based attention detection identifies a primary area of interest (AOI) in the picture that the child is attending to. Second, the system selects a related secondary AOI using relevance distance. Third, crops of the primary and secondary AOIs, together with the full picture and a storytelling prompt, are provided as input to the LLM. Fourth, the LLM generates spoken narration that starts from the primary AOI, connects it to the secondary AOI, and ends with an open-ended question.}
\label{fig:pipeline}
\end{figure}

\zekun{To realize these goals, Ollie’s pipeline consists of four steps, as shown in~\autoref{fig:pipeline}: (1) detecting the child’s attentional state from gaze behavior, that is, whether the child is attending to a specific image area (called the primary AOI) or whether they are exploring the image, (2) selecting a secondary AOI that is relevant to the child’s current visual focus but has not yet been talked about, and (3) generating a short narrative that connects the current primary AOI and the selected secondary AOI. The system then outputs the narration via text-to-speech (4). The implementation of steps (1) - (3) are described in the following. }


\subsection{Attentional State Detection from Gaze}
\label{sec:gaze-attention}

\zekun{Ollie first uses gaze to decide whether the child is clearly attending to one AOI. We model this as a two-state Hidden Markov Model (HMM), with states corresponding to \emph{attention} and \emph{no-attention}. The \emph{attention} state means that the child’s gaze is mainly concentrated on one dominant AOI, which Ollie can use as the starting point for narration. The \emph{no-attention} state means that gaze is dispersed, unstable, or not clearly linked to any single AOI.}

At each time step~$t$, corresponding to a 500\,ms window, the model observes a gaze-derived feature $x_t$ that captures how strongly the child's gaze is focused on a single AOI. 

Each hidden state is associated with a Gaussian Mixture Model (GMM) emission distribution with two components,
\[
p(x_t \mid z_t = i) = \sum_{m=1}^{2} w_{im}\,\mathcal{N}(x_t \mid \mu_{im}, \Sigma_{im}),
\]
where $w_{im}$, $\mu_{im}$, and $\Sigma_{im}$ denote the mixture weight, mean, and covariance of component~$m$ under state~$i$, respectively.

To compute the gaze feature, $x_t$, used as the observation in the HMM, we extend standard AOI-based gaze measures \cite{hessels2016area} by defining a dynamic, area-normalized fixation time. Specifically, for each AOI on the picture, we compute the fixation time, defined as the sum of fixation durations that fall inside that AOI within the time window (500ms). Then, to account for different sized AOIs, all fixation time values are normalized by the AOI’s pixel coverage \cite{maslowska2020consumers,lenski2023emotional}. Intuitively, a high value of $x_t$ indicates that \zekun{an attention state, where the child’s gaze is concentrated on one dominant AOI. A low value indicates a no-attention state, where gaze is dispersed across AOIs or not clearly concentrated on any single region.}

Next, we identify the dominant AOI in that window—the AOI with the highest fixation time. We then compute the \textbf{Fixation Ratio on the Dominant AOI (FR-D)} as the proportion of fixation time on the dominant AOI divided by the total fixation time across all AOIs in that window. This ratio reflects how strongly the child's gaze is concentrated on a single AOI during the time window—i.e., how likely the dominant AOI captured and held the child's visual attention, as indicated in Equation~\ref{eq:frd}.

\begin{equation}
\text{FR-D}_t =
\frac{\tilde{F}_{\text{dominant},t} + 1}
     {\sum_{i \in \mathcal{A}} \tilde{F}_{i,t} + 2}
\label{eq:frd}
\end{equation}
where $\tilde{F}_{i,t}$ is the area-normalized fixation time for AOI $i$ in window $t$. The constants 1 and 2 are Laplace smoothing terms to avoid extreme values near~0 or~1. The resulting feature value $x_t = \text{FR-D}_t \in (0,1)$ serves as the observation at time~$t$.

To ensure stable parameter initialization, we include a 5\,s \textit{warm-start} phase (10 consecutive time steps) during which no state inference is performed. 
A batch EM algorithm is applied to the warm-start data to estimate the initial parameters $\Theta_0 = \{\pi, A, w, \mu, \Sigma\}$, 
where $\pi$ denotes the initial state distribution and $A$ the transition matrix. 
After initialization, the system transitions to an \textit{online EM} procedure: 
at each subsequent 500\,ms interval, the model (i) updates the filtered posterior probabilities $P(z_t \mid x_{1:t})$ via the forward algorithm, 
(ii) computes local expected sufficient statistics for transitions and emissions, and 
(iii) performs incremental parameter updates using exponential moving averages. 
Different update rates are applied to ensure stability and responsiveness: 0.1 for emission parameters and 0.05 for transition and mixture parameters.
 
\subsection{Gaze-Informed AOI Selection}
\label{sec:gaze-selection}

Once Ollie determines that assistance should be provided, it generates gaze-informed narrative assistance that is grounded in the child's current visual focus and extends the narration toward related, previously unexplored parts of the picture. The core idea is to use the gaze-based attention detection method described in \autoref{sec:gaze-attention} to identify the primary AOI that currently captures the child’s visual attention, and then proactively introduce a second, contextually relevant but unassisted AOI to encourage further exploration. Specifically, the selection system integrates two components:



\paragraph{AOI representation.} Each AOI is represented using three complementary feature types: 
(1) a list of objects contained within the region (semantic features); 
(2) its on-screen coordinates (spatial features); and 
(3) a gaze-time log indicating when the region was last viewed (temporal features).
These representations support evaluation of how each AOI relates to the child's current visual focus.

\paragraph{Relevance distance estimation.}
For each AOI, the system computes three distances relative to the currently fixated AOI:
semantic distance (cosine distance over TF--IDF vectors of object lists),
spatial distance (Euclidean distance between AOI centers), and 
temporal distance (time elapsed since the AOI was last fixated). 
Distances are normalized and combined into a weighted one:

\begin{equation}
D = 0.2D_{\text{semantic}} + 0.4D_{\text{spatial}} + 0.4D_{\text{temporal}}.
\label{eq:relevance_distance}
\end{equation}
The AOI with the lowest distance score among those that have not yet received storytelling assistance (the \textit{unassisted} set) is selected as the secondary AOI to be described.

\subsection{Narrative Generation}
\label{sec:narrative_generation}

\begin{figure}[t]
\centering
\includegraphics[width=6.0in]{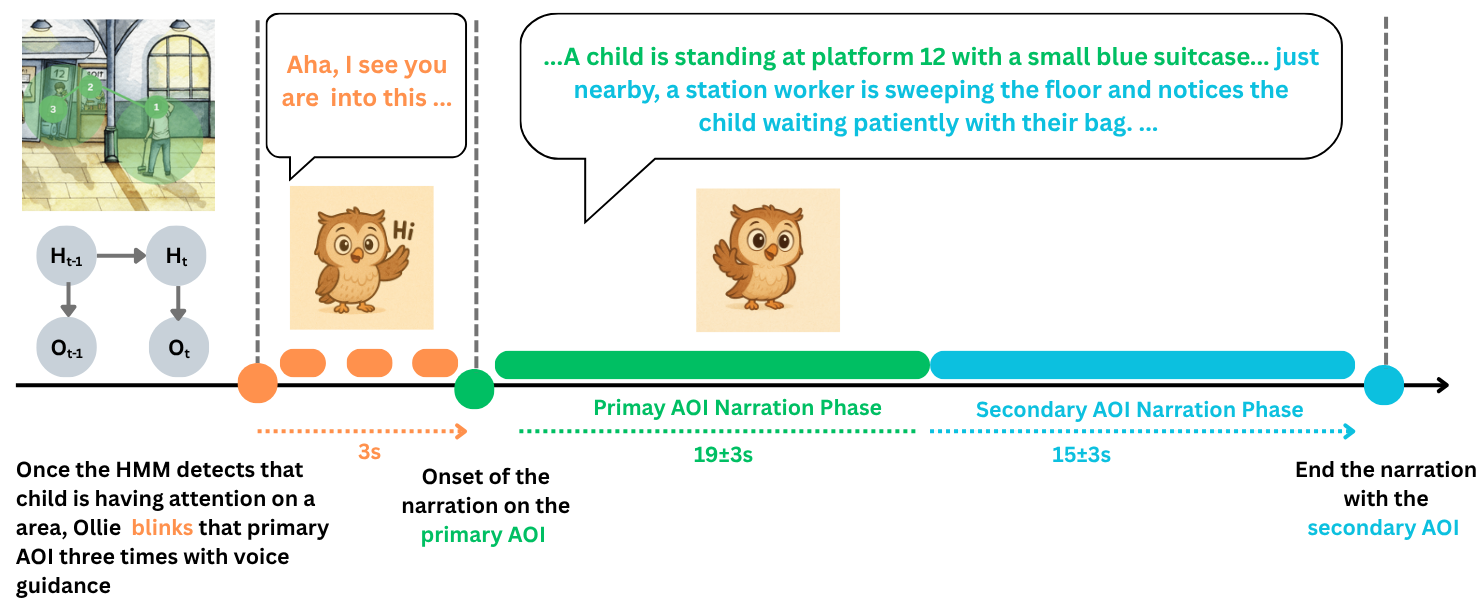}
\caption{A complete timeline for an assistance instance given by Ollie.}
\label{fig:timeline}
\end{figure}

\zekun{As shown in \autoref{fig:timeline}, once the HMM described in \autoref{sec:gaze-attention} detects that the child is attending to a specific area, Ollie highlights the corresponding AOI in orange with three brief blinks over a period of 3 seconds while providing a short voice prompt. In addition to guiding the child’s attention, this blinking phase provides time for the system to generate and buffer the LLM-based narrative assistance.} Specifically, we generate narrative assistance using an LLM with three sources of visual context: the currently gazed \emph{primary AOI} (Section~\ref{sec:gaze-attention}), the selected \emph{secondary AOI} (Section~\ref{sec:gaze-selection}), and the full image. 

The LLM is instructed to treat the task as a story-chaining process, in which each new narration builds on the previously generated content. For the first assistance instance, we provide the model with a short, predefined \emph{mini-context} that specifies basic situational information (e.g., place or time). The model then generates a short narrative that (1) describes the main characters or events in the primary AOI, (2) introduces the secondary AOI using a simple spatial or contextual bridge (e.g., “nearby” or “not far away”), and (3) connects the two regions by describing a plausible relationship or interaction. To encourage factual consistency, we also provide the model with lists of pre-detected objects in both AOIs for cross-checking. The narration ends with a brief open-ended hook (e.g., a question) grounded in the secondary AOI to spark the child’s curiosity.

This hook is also used to connect subsequent assistance instances where the model first briefly addresses the hook from the previous narration and then repeats the same structure: describing the current primary AOI, introducing a new secondary AOI, and ending with a new hook to encourage continued exploration. In the user study, Ollie generated and delivered all spoken narrations in German, which was the language used by participating children. In this paper, we translate example narrations and prompt excerpts into English for readability. The complete prompt we used in the study is presented in the appendix.

\section{Offline Validation of Gaze-Informed Assistance Models}
\label{validation}
\zekun{We first conducted an offline validation using gaze data collected while children freely explored the pictures. The purpose was to assess whether Ollie’s modeling approach reflected the design goals described in \autoref{sec:design-goals} by aligning assistance with children’s moment-to-moment attention and ongoing exploration. Specifically, we evaluated whether the HMM triggered at moments when children were attending to a specific AOI (DG1), and whether the relevance-based selection of the secondary AOI is aligned with where children looked at next (DG2).}

\subsection{Validation Dataset and Attention Labels}
\label{Labelling}

\zekun{Because there is no single operational definition of when a child is meaningfully attending to a specific AOI during free exploration, we evaluated Ollie’s timing decisions using two proxy-labeling approaches, which we describe below. We collected free-exploration gaze data from four children across nine images with each lasting at least 30 seconds. We replayed these recordings as a sequence of candidate assistance intervals, using the same 500 ms window length as Ollie’s online decision process.}

\zekun{The first approach used expert-annotated AOI-level attention. Two experts in eye tracking research and authors of this paper independently annotated the recorded gaze data by reviewed replay videos of the child’s picture exploration. Each video showed the picture seen by the child, the predefined AOI segmentation, and moving gaze heatmaps (aggregated over 500ms windows) indicating where the child was looking over time. Experts marked moments when the child appeared to attend to a specific AOI (rather than broadly exploring the picture). These marked moments were then mapped to the corresponding 500 ms intervals in the child’s gaze data log. Marked intervals were labeled as attention, with the annotated AOI treated as the primary AOI; all other intervals were labeled as no-attention. The experts agreed more often when attention to an AOI was clear and sustained, while shorter fixations were harder to get consistent labels. After retaining only agreed-upon intervals, the final dataset contained 669 attention and 285 no-attention intervals.}

\zekun{The second approach was to use future-gaze stability as a rule-based behavioral proxy. We selected this approach to reduce reliance on subjective judgments for intervals containing shorter fixations. For each candidate interval, we identified the dominant AOI (i.e. the AOI with the highest fixation duration in the current 500ms window). If the same AOI remained dominant in the following 500 ms interval, we labeled the current interval as attention, treating that AOI as the primary AOI. Otherwise, the interval was labeled as no-attention. This proxy reflects whether the current AOI continued to serve as a stable visual anchor shortly after the candidate trigger moment, and produced 711 attention instances and 607 no-attention instances.}

\subsection{Timing Validation}

\zekun{As shown in \autoref{tab:timing-validation}, we compared the HMM timing model introduced in \autoref{sec:gaze-attention} with two simple baselines: a fixed 3s trigger and a 3s dwell trigger. The fixed baseline triggered assistance every 3 seconds, while the dwell baseline triggered only after the child had explored the same AOI for 3 seconds. Across both proxy-labeling approaches, the HMM achieved the highest F1 score, mainly due to substantially higher recall than the two baselines. The dwell baseline achieved the highest precision, but missed many candidate attention intervals, indicating that it was more conservative and less suitable for timely proactive assistance.}

\begin{table}[h]
\centering
\begin{tabular}{llccc}
\toprule
\textbf{Proxy label} & \textbf{Method} & \textbf{Precision} & \textbf{Recall} & \textbf{F1} \\
\midrule
\multirow{3}{*}{Expert-annotated}
& Fixed       & 0.669 & 0.166 & 0.266 \\
& 3s dwelling & \textbf{0.989} & 0.262 & 0.414  \\
& HMM         & 0.793 & \textbf{0.774} & \textbf{0.784}\\
\midrule
\multirow{3}{*}{Future-gaze}
& Fixed       & 0.570 & 0.179 & 0.272 \\
& 3s dwelling & \textbf{0.761} & 0.193 & 0.308 \\
& HMM         & 0.595 & \textbf{0.720} & \textbf{0.651} \\
\bottomrule
\end{tabular}
\caption{Timing validation results under two proxy-labeling approaches: expert annotated and future gaze-based AOI-level attention.}
\label{tab:timing-validation}
\end{table}

\subsection{AOI-Selection Validation}

\zekun{Based on the annotated AOI for each attention interval described in \autoref{Labelling}, we defined the secondary-AOI proxy label as the next different AOI attended afterward. We then evaluate whether the relevance-based selection introduced in \autoref{sec:aoi-selection} picks AOIs that lie along the child's ongoing exploration. Concretely, we compute the accuracy in predicting  the next-AOI on the child's own path through the picture. Each image contains 12 to 15 candidate AOIs, and where a child looks next is only partly determined at any given moment. Accuracy for a well-aligned selection method is therefore expected to remain low in absolute terms.}
\zekun{We compared the combined relevance approach with four alternatives: a random method that selected any AOI other than the currently attended one, and three single-distance methods that selected the nearest AOI based only on semantic, spatial, or temporal distance, as introduced in \autoref{sec:gaze-selection}. As indicated in \autoref{tab:aoi-selection-validation}, the combined relevance approach achieved the highest accuracy under both proxy-labeling approaches. This suggests that considering all three dimensions together better aligns secondary-AOI selection with children’s natural attention transitions than relying on any single dimension or random selection.}

\begin{table}[h]
    \centering
    \begin{tabular}{lcc}
        \toprule
        \textbf{Method} & \multicolumn{2}{c}{\textbf{Accuracy}} \\
        \cmidrule(lr){2-3}
        & \textbf{Expert-annotated} & \textbf{Future-gaze} \\
        \midrule
                Random
        & 0.098 & 0.101 \\
        Semantic Only
        & 0.191 & 0.167 \\
                Spatial Only
        & 0.129 & 0.171 \\
        Temporal Only
        & 0.180 & 0.202 \\

        Combined Relevance
        $(0.2D_{\mathrm{sem}} + 0.4D_{\mathrm{spa}} + 0.4D_{\mathrm{temp}})$
        & \textbf{0.209} & \textbf{0.215} \\
        \bottomrule
    \end{tabular}
    \caption{AOI-selection validation results under the expert-annotated and future-gaze proxy labels. Accuracy indicates the proportion of instances in which the selected secondary AOI matched the AOI the child attended to next.}
    \label{tab:aoi-selection-validation}
\end{table}


\section{User Study}
\zekun{After validating that the timing and AOI-selection components of our modeling approach aligned with children’s natural gaze patterns (\autoref{validation} ), we evaluated Ollie in a live picture-exploration setting, where children received real-time narrative assistance about the pictures. We conducted a within-subject experiment to examine how Ollie’s gaze-informed assistance further shaped children’s visual attention on picture and how children and their parents or guardians experienced such assistance. Specifically, we addressed the following research questions (RQs):}

\textbf{RQ1:} \zekun{How does grounding proactive AI assistance in children’s gaze behavior affect their visual attention during picture exploration?}

\textbf{RQ2:} \zekun{How do children and their parents or guardians perceive and evaluate gaze-informed assistance in terms of its usefulness, and their overall experience during picture exploration?}

\zekun{To address the above RQs, we introduced a baseline condition in which Ollie provided the same form of narrative assistance without using children’s gaze information. In this condition, assistance was delivered at fixed three-second intervals, and the primary and secondary AOIs discussed in each narration were selected randomly.}




\subsection{Pilot Study}
\zekun{We began with a pilot study to finalize several aspects of the study setup and Ollie’s assistance behavior, including how long children explored each image, how often and how many times Ollie provided assistance, and which images were used. We recruited and tested the proactive assistant system with ten children aged four to ten years from the local community. The pilot study ran on a desktop computer with a 27-inch screen, to which a Tobii Pro Fusion eye tracker was attached. In the pilot, each child viewed between 6 and 15 picturebook-style images created by different artists, each presenting a visually-rich scene containing multiple characters, objects, and ongoing events. For each image, we asked children whether they found it interesting to explore and suitable for storytelling. We then selected the images that were most consistently perceived as interesting and storytelling-friendly across participants.}

\zekun{Alongside selecting the image materials, we used the pilot study to understand how children naturally paced their interaction with Ollie. The pilot was intentionally open-ended: children could explore as many images as they wished, spend as long as they wanted on each image, and receive as many narrative descriptions as they wanted. Most children chose to move on from an image after approximately three minutes, and their engagement often appeared to decline after around three assistance instances. Children’s feedback also suggested that assistance delivered too frequently would interrupt their self-directed exploration.}

\zekun{These observations informed the procedure of the main study. Ollie initially provided three assistance instances for each image, after which children could decide whether to continue exploring or move to the next image, with a maximum exploration time of five minutes per image. We also introduced a minimum three-second interval between successive assistance triggers to prevent the system from intervening too frequently. Finally, we selected six images (three for each assistance condition) to keep the total picture-exploration duration within approximately 40 minutes and the complete study session within one hour.}

\subsection{Participants}
For the main study, we recruited 26 children through mailing lists, physical flyers, and word of mouth, and conducted the study in two settings: a laboratory and a local kindergarten. \edit{Prior to scheduling sessions, we emailed parents to verify that their child already engaged in at least one picture‑book reading activity per week, either independently or with a parent. This simple check ensured that participants were familiar with picture‑book exploration.} Of the 26 children, two declined to finish exploring all images with Ollie. Also, one session was terminated due to an unrecoverable technical issue, and one child had to leave early when a parent arrived for pickup, before completing the study.

The final sample consisted of 22 children (12 female, 10 male): Among them, we had 11 five-year-olds, 4 six-year-olds, 4 seven-year-olds, and 3 eight-year-olds. Of the final sample, 13 children read picture books once per week, whereas the remaining 9 did so more than once per week.

In addition, 16 parents participated in the study and remained present with their child throughout the sessions conducted in the laboratory setting. Among them, three parents each brought two children and therefore attended two consecutive experimental sessions. One kindergarten teacher also participated in the study during the kindergarten sessions. No parents declined to participate in the interview. The study had been approved by the ethics review board of the authors' affiliated university before data collection.

\subsection{Materials}
\begin{figure}[htbp]
    \centering
    \begin{subfigure}{0.49\textwidth}
        \includegraphics[width=\linewidth]{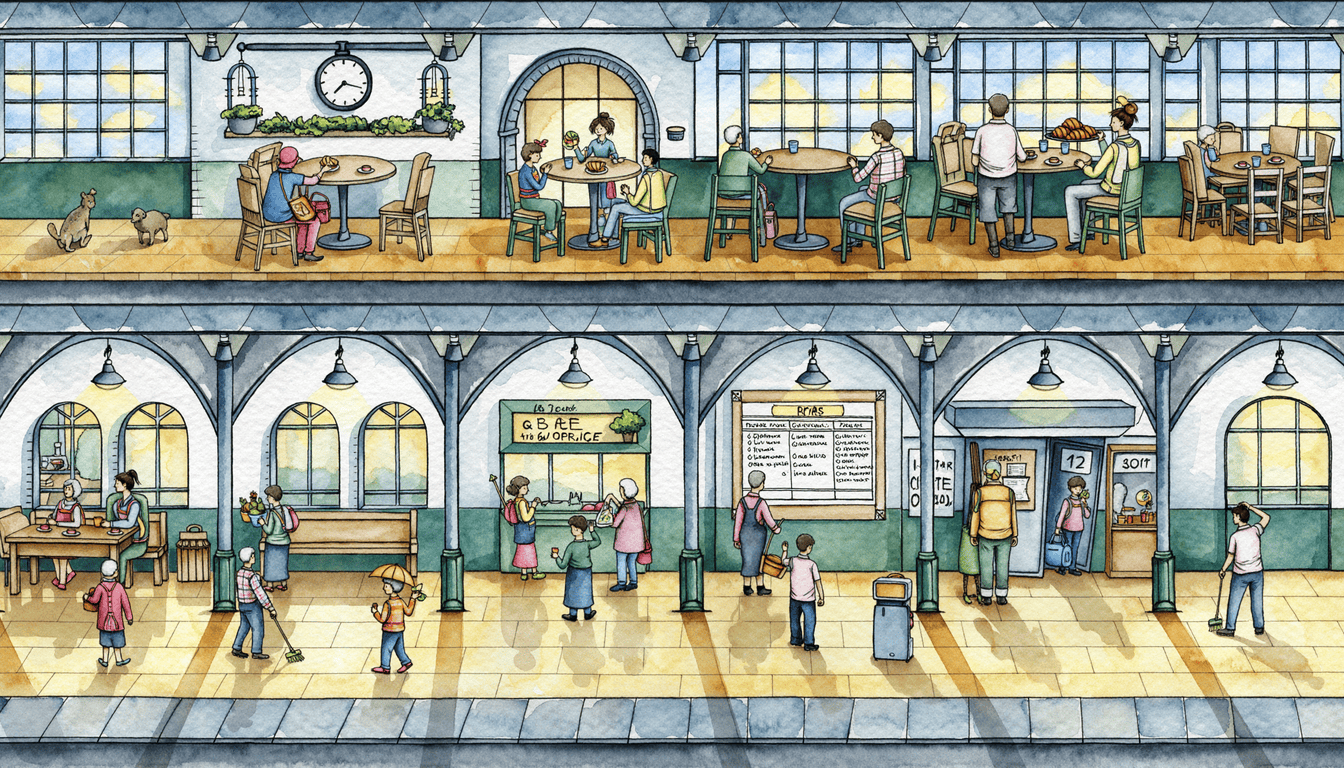}
        \caption{Image used in the Experiment}
        \label{fig:image}
    \end{subfigure}
    \hfill
    \begin{subfigure}{0.49\textwidth}
        \includegraphics[width=\linewidth]{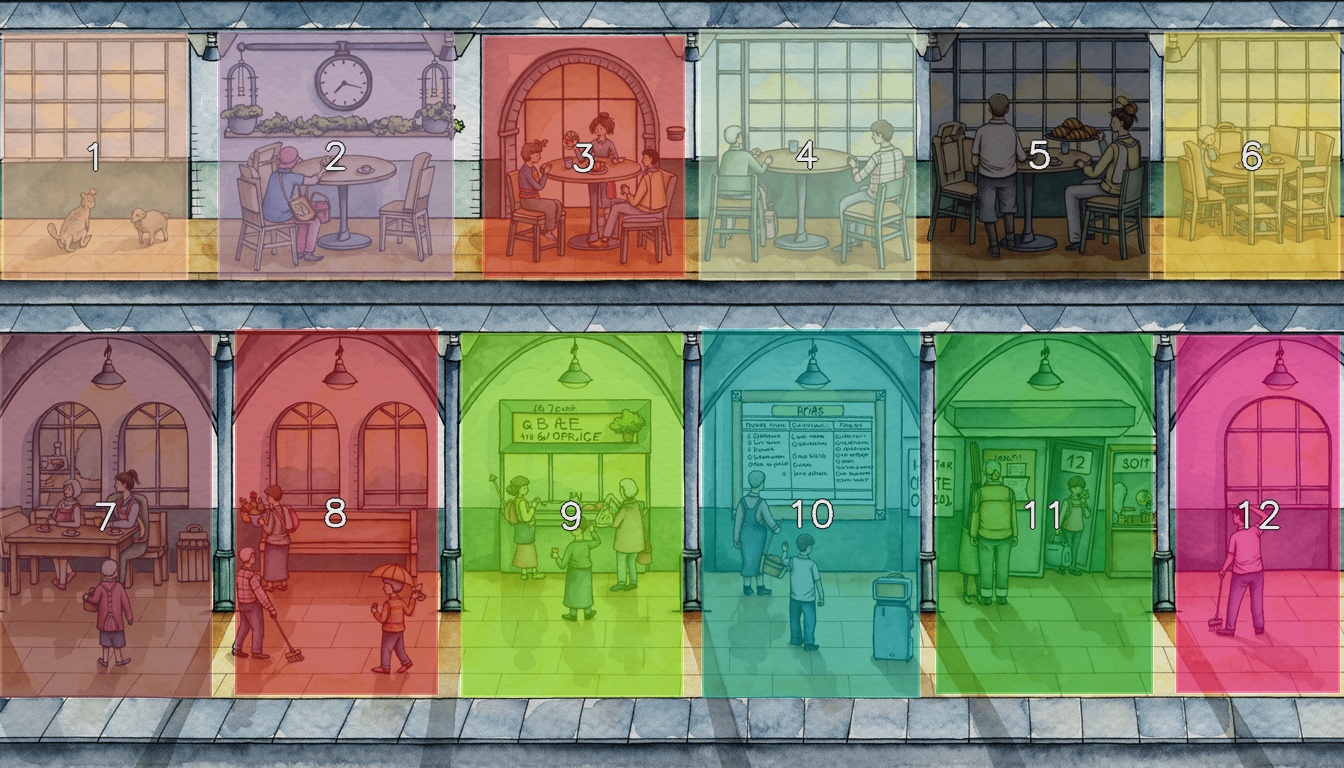}
        \caption{AOIs on the Image}
        \label{fig:aois}
    \end{subfigure}
    \caption{Example picture and its area-of-interest (AOI) segmentation. The picture is divided into semantically meaningful regions used for gaze-based attention detection and narrative assistance.}
    \Description{Two side-by-side views of the same illustrated train-station scene. The left panel shows the original picture, containing an upper and lower station level with people, furniture, doors, windows, and other objects. The right panel overlays the same picture with 12 semi-transparent, color-coded areas of interest, numbered 1 through 12. The AOIs partition the scene into semantically meaningful spatial regions that are used to associate children's gaze with particular parts of the picture.}
    \label{fig:image_examples}
\end{figure}

\edit{For the main user study, we prepared six images in a visually rich 
\textit{Wimmelbook}-style picturebook format.\footnote{Wimmelbooks, 
or \textit{Wimmelbilderbücher}, are picture books characterized by 
dense, full-spread scenes containing many people, animals, objects, 
and small events. See: \url{https://en.wikipedia.org/wiki/Wimmelbilderbuch}.} } The images were selected from three picture books by three different creators, covering three themes: construction sites, humans and animals, and emergency incidents, each having two scenes. In addition, we used AI to generate one image that approximates the visual style and composition of the study materials (\autoref{fig:image}) for the pilot study and for illustration in the paper (as other images are copyrighted). \edit{The six images were organized into three theme-matched pairs. For each participant, one image from each theme-matched pair was assigned to the gaze-based condition and the other to the baseline condition, resulting in three images per condition. This assignment was randomized across participants so that each condition included one image from each theme.}


For each image, AOIs were manually defined to segment the scene into meaningful regions based on interior and exterior spaces, as shown in \autoref{fig:aois}. 
To maintain comparable visual complexity across images, each AOI contained two to five key objects, and each image has a total of 12 to 15 AOIs. 

\subsection{Assistance Conditions}
\label{sec:assistance-conditions}
We employed two assistance conditions in this study: (1) an LLM baseline assistance condition and (2) a gaze-based assistance condition.
In both conditions, the same prompt described in \autoref{sec:narrative_generation} was used to generate spoken assistance with a large language model (GPT-4o). The generated responses were delivered to the child via text-to-speech using the Azure service. Each assistance instance followed the same temporal structure, similar to what is visualized in \autoref{fig:timeline}~: the system first highlighted the \emph{primary AOI} (the starting region of the narration) for 3 seconds, it then started the narration, which first focuses on the primary AOI and then transitions to a \emph{secondary AOI}.

The two conditions differed in their timing and  in how these two AOIs were selected. In the gaze-based assistance condition, the primary AOI and the time of highlighting it was determined based on the child’s current visual attention using the attentional state detection method described in \autoref{sec:gaze-attention}. The secondary AOI was then selected using the relevance estimation procedure described in \autoref{sec:gaze-selection}. In the baseline condition, the highlight started after a fixed 3s interval and both the primary AOI and the secondary AOI were selected at random from the set of available AOIs in the picture. All other features were kept the same.

Note that the primary AOI was visually highlighted for 3 seconds in both conditions before the narration started. This supported children's understanding of what the narration referred to and, importantly for the analysis, directed their gaze to the primary AOI at the start of the narration regardless of how that AOI had been selected, ensuring that the visual attention metrics defined in \autoref{sec:visual-attention} measure what happens during the narration rather than differences in initial orientation. The secondary AOI was introduced only through speech and was not highlighted, so that any attention shift toward it reflects the quality of the verbal suggestion rather than an explicit visual cue.

\subsection{Procedure}
The study was conducted in two settings: a laboratory environment and a local kindergarten. In the laboratory setting, children participated together with their parent/guardian, who remained present during the session. In the kindergarten setting, sessions were conducted in the closed library with a kindergarten teacher present with each participating child. In both cases, the same experimental setup and instructions were used across settings. Specifically, we used a 27 inch desktop monitor screen with a Tobii Pro Fusion eye tracker operating at 250~Hz to collect data. Each session began with a 5-minute introduction in which we explained the task to the child and demonstrated how the AI assistant, Ollie, would provide spoken guidance during the picture exploration. We also explicitly explained that Ollie would use two different assistance strategies (with or without gaze), and that we would need their help in evaluating the system by sharing their thoughts and feedback afterwards. During the study, we checked whether children noticed this difference, including whether Ollie seemed to follow their gaze or point out regions on its own.

Each child explored six picturebook–style images in total. The images were organized into two blocks by assistance condition: three images in the baseline condition and three in gaze-informed assistance condition. Within each block, three images were presented consecutively, with one image randomly selected from each of the three theme-matched pairs. The order of the two blocks was counterbalanced across participants. For each image, Ollie provided three consecutive instances of assistance. After these, we paused the assistance and asked the child whether they wanted to continue exploring the same image or move on to the next one. If the child chose to continue, exploration resumed and Ollie provided another assistance, after which the child was again asked whether they wished to continue or switch to the next picture. This loop continued until the child chose to stop or a maximum time limit of five minutes was reached.

After completing all six images, we conducted a short post-study interview with the child and the accompanying adult (parent/guardian/teacher) separately. We provide interview questions in the following section.

\subsection{Measures}
To address our research questions, we evaluated the proactive assistance from three aspects: (1) \textit{visual attention during exploration} for \textbf{RQ1}, and (2) \textit{subjective feedback from post-study interviews} for \textbf{RQ2}.

\subsubsection{Visual Attention}
\label{sec:visual-attention}

\edit{To explore how Ollie's assistance affected children's visual attention, we analyzed their gaze toward the primary and secondary AOIs referenced in each narration. 
For each assistance instance, we defined a temporal analysis window from the onset of the narration until the narration ends. Note that this window starts after the highlighting period excluding fixations accumulated before the narration started, which in the gaze-based condition include the attention that triggered it. We divide this window into a primary-AOI interval and a secondary-AOI interval (marked in green and blue in~\autoref{fig:timeline}). For each interval, we computed three gaze metrics for the corresponding assisted AOI: (1) \emph{follow-up fixation rate} as the proportion of assistances during which the child fixated the AOI; (2) \emph{fixation duration} as the total time spent fixating that AOI within the interval; and \emph{time to first fixation}, the latency from the onset of the AOI’s verbal mention to the first fixation on it. We further tracked the dominant AOI across the full assistance window, following the definition in \autoref{sec:gaze-attention}, to examine how children’s attention shifted over time on the narrated AOIs compared to other regions of the image.}

\subsubsection{Subjective Feedback}
After completing all six images, each child and the accompanying adult participated in a short post-study interview. The interview focused on three aspects: (1) participants’ overall experience with Ollie, including whether they liked the stories and wanted to use such an assistant again; (2) their preference between the baseline and gaze-based assistance conditions, together with any underlying reasons; and (3) the perceived usefulness of such an assistant in everyday settings, such as home reading or kindergarten classroom use, as well as possible improvements to the system.

For children, we asked about their impressions of Ollie, which stories or images they liked the most, whether they preferred one assistance strategy over the other, and whether they would like to use Ollie during picture exploration at home. 
We also asked how Ollie’s storytelling compared with the support they usually receive.

For accompanying adults, the interview focused on whether they would consider using such an assistant at home or in school, which assistance strategy they preferred, how Ollie compared with parent or teacher storytelling, and what changes would make the system more helpful in practice.

The full interview questions are provided in the Appendix.
\section{Results}
In the following, we first describe our data pre-processing steps. We then report on the impact of Ollie’s gaze-based proactive assistance on two aspects of children’s picture reading: visual attention (RQ1), and subjective feedback (RQ2). Specifically, we aggregated the gaze data at the participant level to account for repeated measurements within each child. After checking for normality with the Shapiro-Wilk Test, we performed statistical hypothesis testing, either a paired \textit{t}-test (in case of normality of continuous variables) or the Wilcoxon signed-rank Test (in case of non-normality or non-continuous variables).

\subsection{Gaze Data collection and Pre-Processing}
\label{sec:data_processing}
For gaze analysis, we treated the data as time-series gaze records corresponding to individual Ollie assistance instances. In total, we had 405 gaze data (assistance instance) from 22 participants with 203 in gaze-based assistant condition, and 202 in baseline condition. Each gaze point was captured as an (x,y) screen coordinate, together with their corresponding timestamp. From the raw gaze points, fixations were identified based on specific criteria: low dispersion (25\,px) and adequate duration (50\,ms), using fixation detection function from the PyGaze package~\cite{dalmaijer2014pygaze}. To remove data with poor tracking quality, we used the highlighted primary AOI shown during each assistance instance as a validation check. As described in the assistance conditions (\autoref{sec:assistance-conditions}), the primary AOI was visually highlighted while the assistance was delivered. Following prior work ~\cite{feit2017toward}, we used the highlighted primary AOI as a validation check. For each assistance instance, we examined whether the participant made at least one fixation on the highlighted AOI. Instances with no such fixation were excluded from the gaze analysis. If this occurred in more than half of the assistance instances for a given image, we excluded the entire image and its corresponding gaze data. If more than three images were excluded for a participant, we excluded the participant entirely, as this would eliminate all data from one assistance condition. After this filtering procedure, one participant was excluded, and 13.2\% of the gaze data were removed.

\subsection{Analysis of Visual Attention}
\label{sec:analysis_visual_attention}

\begin{figure}[t]
    \centering
    \begin{subfigure}[t]{0.49\textwidth}
        \centering
        \includegraphics[width=\linewidth]{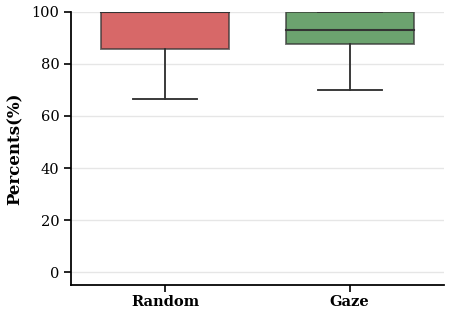}
        \caption{Follow-up Fixation Rate on Primary AOI.}
        \label{fig:fixation_rate_primary}
    \end{subfigure}
    \hfill
    \begin{subfigure}[t]{0.49\textwidth}
        \centering
        \includegraphics[width=\linewidth]{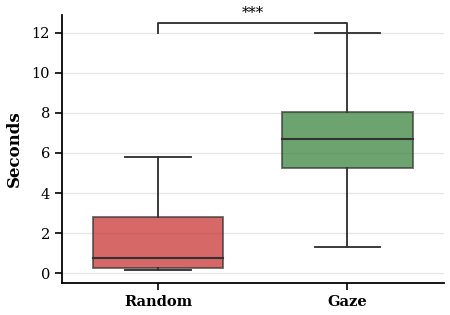}
        \caption{Fixation Duration on Primary AOI.}
        \label{fig:fixation_duration_primary}
    \end{subfigure}
    \caption{Follow-up fixation rate and fixation duration on primary AOI across conditions.}
    \Description{Two boxplots compare children's gaze behavior on the primary AOI between the random baseline and gaze-informed conditions. The left plot shows follow-up fixation rate in percent; both conditions have similarly high fixation rates, with substantial overlap between their distributions. The right plot shows fixation duration in seconds; fixation durations are substantially longer in the gaze-informed condition than in the random baseline, with a statistically significant difference indicated by three asterisks.}
    \label{fig:primary_aoi}
\end{figure}

As described in \autoref{sec:visual-attention}, we first examined fixations on the two assisted AOIs separately. For the primary AOI, both conditions yielded similarly high follow-up fixation rates (Figure~\ref{fig:fixation_rate_primary}), likely because the primary AOI was visually highlighted in both conditions: the baseline condition achieved \(M = 91.66\%\), \(SD = 10.27\%\), and the gaze-based condition achieved \(M = 90.81\%\), \(SD = 10.73\%\), with no significant difference between the conditions (\(p > .05\)). In contrast, fixation duration on the primary AOI differed substantially across conditions (Figure~\ref{fig:fixation_duration_primary}). Children with the gaze-based assistance spent significantly longer fixating on the primary AOI (\(M = 6.48\) s, \(SD = 2.53\)) than with the baseline condition (\(M = 1.67\) s, \(SD = 0.75\); \(p < .001\)). Because the analysis window begins only at narration onset, this difference does not include the fixation time that preceded the trigger in the gaze-based condition. Together with the equivalent follow-up fixation rates, this indicates  that children were equally likely to attend to the narrated region in both conditions, but stayed with it longer when it was the region they had already been attending to before the assistance started.

\begin{figure}[t]
    \centering
    \begin{subfigure}[t]{0.49\textwidth}
        \centering
        \includegraphics[width=\linewidth]{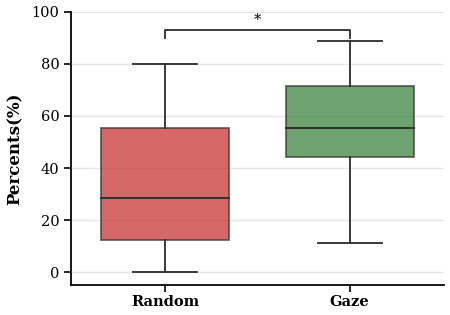}
        \caption{Follow-up Fixation Rate on Secondary AOI.}
        \label{fig:fixation_rate}
    \end{subfigure}
    \hfill
    \begin{subfigure}[t]{0.49\textwidth}
        \centering
        \includegraphics[width=\linewidth]{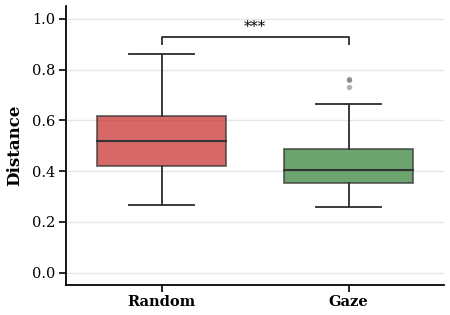}
        \caption{Relevance Distance}
        \label{fig:relevance}
    \end{subfigure}
    \caption{Follow-up fixation rate on secondary AOI and relevance distance between primary \& secondary AOIs across conditions.}
    \label{fig:relevance}
\end{figure}

Next, we examined the secondary AOI, which was selected based on the relevance measure and introduced only through spoken assistance, without any visual highlighting. As indicated in \autoref{fig:fixation_rate}, the results show that the gaze-based condition yields a significantly higher follow-up fixation rate on the secondary AOI (M = 54.06\%, SD = 20.92\%) than the baseline condition (M = 35.48\%, SD = 28.34\%), with a statistically significant difference ($p < .05$). To directly examine how the two selection strategies differed, we compared the relevance distance between the primary AOI and the selected secondary AOI in each condition, as defined in Equation~\ref{eq:relevance_distance}. In the gaze-based condition, relevance distance between the primary AOI and the secondary one (M = 0.43, SD = 0.10) is significantly lower than the baseline condition (M = 0.52, SD = 0.13, p < .001). These results suggest that the relevance distance-based selection helped children follow Ollie’s spoken guidance: by choosing a secondary AOI that was closer to the child’s current focus, the gaze-based condition made the next narrated region easier for children to attend to.

\begin{figure}[t]
    \centering
    \begin{subfigure}[t]{0.49\textwidth}
        \centering
        \includegraphics[width=\linewidth]{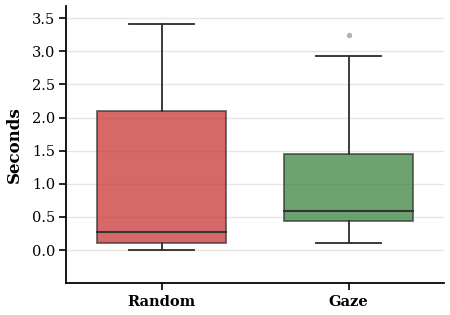}
        \caption{Fixation Duration on Secondary AOI.}
        \label{fig:fix_duration}
    \end{subfigure}
    \hfill
    \begin{subfigure}[t]{0.49\textwidth}
        \centering
        \includegraphics[width=\linewidth]{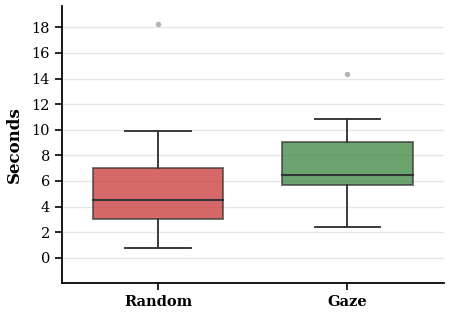}
        \caption{First Time to Fixation on Secondary AOI.}
        \label{fig:fix_firsttime}
    \end{subfigure}
    \caption{Follow-up fixation duration and first time to fixation on secondary AOI across conditions.}
    \Description{Two boxplots compare children's gaze behavior on the secondary AOI between the random baseline and gaze-informed conditions. The left plot shows fixation duration in seconds; the distributions overlap substantially, with similar median values across conditions. The right plot shows time to first fixation in seconds; children generally took several seconds to first fixate the secondary AOI in both conditions, and the distributions again overlap considerably.}
    \label{fig:fix_duration_firsttime}
\end{figure}

For assistance instances in which children eventually fixated the secondary AOI, we next examined how much attention they allocated to that region. Overall, attention to the secondary AOI remained limited. For fixation duration on the secondary AOI (\autoref{fig:fix_duration}), children spent only a short time looking at the region in both conditions, with a mean fixation duration of \(M = 1.08\) s (\(SD = 0.90\)) in the gaze-based condition and \(M = 1.02\) s (\(SD = 1.13\)) in the baseline condition; this difference was not statistically significant (\(p > .05\)). Even among children who attended to the secondary AOI the most, the average fixation duration did not exceed 3.5 s. We also found that children generally needed substantial time to locate the AOI referenced by Ollie. Time to first fixation was \(M = 6.65\) s (\(SD = 2.26\)) in the gaze-based condition and \(M = 5.65\) s (\(SD = 4.26\)) in the baseline condition, with no significant difference between conditions (\(p > .05\)).

Beyond the aggregate gaze metrics computed over the full assistance window, the temporal distribution of dominant AOI roles provides insight into how children’s attention evolved as Ollie guided them across two AOIs selected under different strategies. Specifically, we conducted a temporal analysis of dominant AOI role based on \(\tilde{F}_{\text{dominant},t}\) in \autoref{eq:frd}. Because assistance duration varied across instances (\(M = 34.05\) s, \(SD = 3.98\) s), we normalized each assistance instance by dividing the primary-AOI narration phase and the secondary-AOI narration phase into five bins each. Within each bin, we identified the dominant AOI by comparing proportional fixation duration across AOIs and classified it as the primary AOI, the secondary AOI, or other AOIs.

\begin{figure}[t]
\centering

\begin{subfigure}{\linewidth}
    \centering
    \includegraphics[width=5.0in]{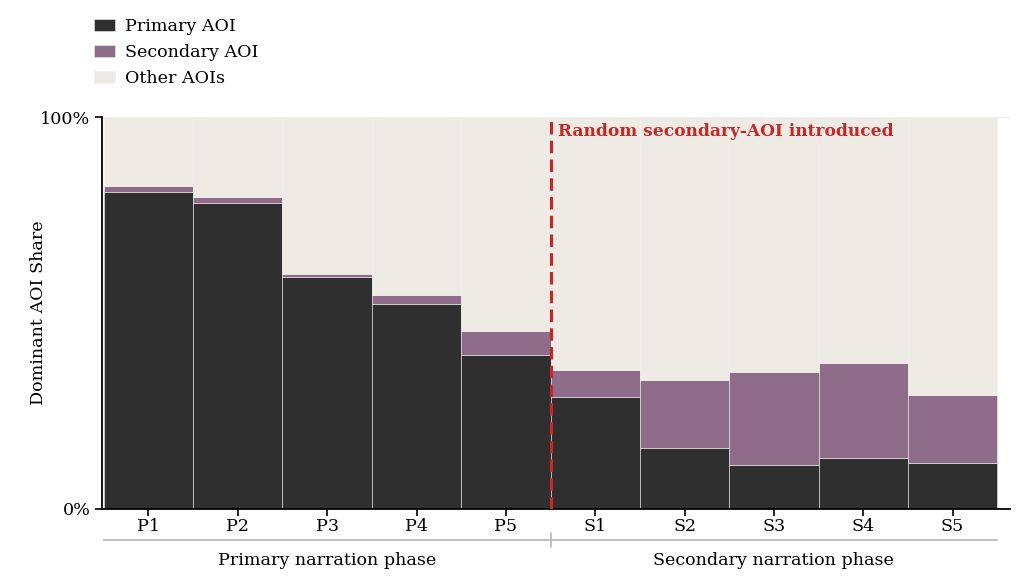}
    \caption{Baseline assistance condition.}
    \label{fig:aoi_shift_random}    
\end{subfigure}
\begin{subfigure}{\linewidth}
    \centering
    \includegraphics[width=5.0in]{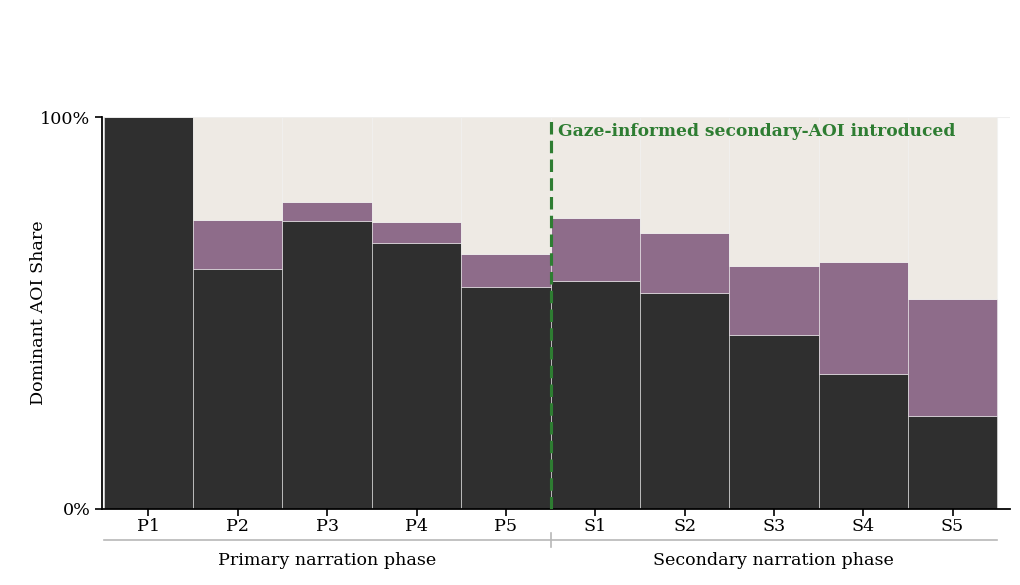}
    \caption{Gaze-based assistance condition.}
    \label{fig:aoi_shift_gaze}
\end{subfigure}

\caption{Temporal shift of dominant AOI role across the full assistance window under the two assistance conditions. For each assistance instance, the narration was divided into a primary-narration phase and a secondary-narration phase, and each phase was normalized into five equal temporal bins. Within each bin, the dominant AOI was classified as the primary AOI, secondary AOI, or other AOIs based on proportional fixation duration. The dashed line marks the onset of the secondary-AOI description.}
\Description{Two stacked bar charts show how children's dominant gaze shifted over ten normalized time bins, from five bins in the primary-narration phase (P1--P5) to five bins in the secondary-narration phase (S1--S5). Dark bars represent the primary AOI, purple bars the secondary AOI, and the remaining share corresponds to other AOIs. A dashed vertical line between P5 and S1 marks when the secondary AOI is introduced. In the baseline condition, dominant attention to the primary AOI decreases sharply across the primary-narration phase and remains low after the secondary AOI is introduced, while attention to the secondary AOI increases. In the gaze-based condition, the primary AOI remains dominant for longer and declines more gradually, while the secondary AOI gains attention progressively during the secondary-narration phase. Overall, the gaze-based condition shows a smoother transition between the two narrated AOIs than the baseline condition.}
\label{fig:aoi_shift}
\end{figure}

As shown in \autoref{fig:aoi_shift}, both conditions exhibited a shift in dominant attention from the primary AOI toward the secondary AOI after the onset of the secondary-AOI description. However, the transition patterns differed. In the gaze-based condition (\autoref{fig:aoi_shift_gaze}), the secondary AOI already accounted for a non-trivial share of dominant attention during the primary narration phase, suggesting that relevance-based selection increased the likelihood that children had already noticed the upcoming region. Moreover, attention to the primary AOI remained relatively sustained into the secondary narration phase, consistent with the longer fixation duration on the primary AOI reported in \autoref{fig:fixation_duration_primary}. By contrast, in the baseline condition (\autoref{fig:aoi_shift_random}), dominant attention to the primary AOI dropped more sharply after the secondary AOI was introduced, accompanied by a rapid increase in attention to the secondary AOI. These patterns suggest that gaze-based assistance supported a smoother and more gradual transition between the two narrated regions, whereas the assistance without gaze elicited a more abrupt reorientation of attention toward the newly mentioned AOI.

\subsection{Analysis of Subjective Feedback}
\label{sec:analysis_subjective_feedback}

We analyzed the post-study interview data using a combination of descriptive coding and exploratory thematic analysis. For questions with clear evaluative or preference-based responses, such as whether participants liked Ollie, which type of assistance they preferred, and whether they would use such a system at home or school, responses were coded into simple categories and summarized using counts. For open-ended responses, one author conducted an iterative thematic analysis to identify recurring reasons related to participants’ preferences, perceived usefulness, and suggestions for improvement. The coding scheme was refined through repeated review of the interview notes to ensure that categories remained consistent across responses. Children, parents, and teacher responses were analyzed separately. We describe high-level themes from this analysis for each of the stakeholders (children, parents, and teachers).

\subsubsection{Children}
\textit{Preference for the gaze-based assistance.} When asked directly which assistance condition they preferred, 15 of 22 children chose the gaze-based condition, 4 chose the baseline condition, and 3 reported no preference. The most common reasons for preferring the gaze-based condition were better alignment with their interests (8 mentions) and a greater sense of control over what Ollie talked about (4 mentions). In contrast, children who preferred the baseline condition referred to its greater surprise and lower sense of constraint, for example not having to keep looking at a particular part of the picture. When asked which version they would want to use at home, this pattern largely remained, as 11 of the 15 children who preferred the gaze-based condition also selected it for home use. However, the preference was weaker in this context, as several children switched to the baseline version or reported being unable to decide, suggesting that preferences expressed during the study did not always translate directly into everyday use intentions.

\textit{Overall experience when being assisted with Ollie was positive across conditions.} 20 of the 22 children reported that they liked Ollie’s descriptions in the gaze-based condition, and 19 of 22 reported liking them in the baseline condition. Similarly, when asked whether there were parts of Ollie’s descriptions that they did not like, most children responded “no” in both conditions (18/22 in the gaze-based condition and 19/22 in the baseline condition), suggesting that there was no clear rejection of Ollie’s assistance. 

\textit{Would-use intention was mostly positive.} Fifteen of the 22 children said that they would want Ollie to accompany them while looking at pictures, whereas 3 said no and 4 were unsure. In explaining this choice, among the 15 children who said they would want Ollie, 4 children mentioned that Ollie added information beyond what was directly visible in the picture, and 3 described the experience as fun. At the same time, 11 of the 22 children were unable to provide a specific reason for their answer, indicating that while the majority of children are willing to use Ollie, they were often not able to articulate exactly why.

\textit{Comparison of the reading experience with parents.} Fourteen of the 22 children reported that they did not usually read picture books with a parent, leaving only 8 children who could make a meaningful comparison between reading with parents and reading with Ollie. Among those 8 children, 3 preferred Ollie’s storytelling, 3 rated Ollie and their parent as comparable, and 2 preferred their parent’s storytelling. This finding suggests no clear pattern of preference.

\subsubsection{Parents}
\textit{Parents strongly preferred the gaze-based assistant.} Fifteen of the 16 parents preferred the gaze-based version over the baseline version. The most common reason was that gaze-based assistance better reflected what the child was currently interested in. Several parents also felt that this version made the child more active in guiding the experience. The only parent who preferred the non-gaze-based version raised privacy and security concerns related to eye tracking.

\textit{Parents’ evaluations of Ollie’s performance were generally positive, although often with limitations around the quality of interactions.} No parent explicitly rejected Ollie’s performance. Instead, most described it positively while also noting limitations, particularly regarding voice quality, interactivity, or linguistic naturalness. This suggests that parents generally saw promise in the system, but also viewed its current form as needing refinement before broader everyday use.

\textit{Parents were divided on whether they would use Ollie or not.} When asked whether they would consider using such an AI assistant in children’s daily reading activities, 7 of the 16 parents said no, 6 said yes, and 3 were uncertain. Compared with the children’s responses, parents were more cautious overall. The most common concerns were screen time and the wish to preserve parent-child interaction during reading. For example, some parents emphasized that children already spend enough time with screens, and others stressed that shared reading should not lose its role as a bonding activity. Parents who responded positively often described Ollie as a useful supplement rather than a replacement, especially when a child is reading alone, when a parent is busy, or as a possible learning support such as for language learning.

\textit{Suggestions for improvement focused mainly on language quality and interactivity.} When asked how the system could be improved, parents most often mentioned problems with language quality, including grammar issues, inappropriate word choice, unnatural pronunciation, delays in story generation, repetitive narration, and a voice that sounded insufficiently natural. A second recurring suggestion was to make Ollie more interactive. Parents proposed that the system should allow children to ask questions, explain unfamiliar words, respond more clearly to children’s input, and generally support a more conversational form of storytelling. Some parents also suggested specific use cases, such as bedtime listening or language-learning activities on a tablet.

\subsubsection{Teachers}
\textit{The kindergarten teacher saw Ollie as a situational classroom support tool.} The teacher suggested that it could be useful during rest periods, with older preschool children, or as part of specific projects, rather than in routine classroom use. Their preference between assistance types depended on the child and the activity: the baseline version could help guide children who fixate only on one part of a picture, whereas the gaze-based version seemed better suited for free exploration. The main limitation she identified was the lack of full interactivity, particularly Ollie’s inability to answer children’s questions. She also suggested possible use cases beyond picture exploration, including support for children with language barriers and for explaining classroom routines.
\section{Discussion}
\subsection{Gaze as an Implicit Signal for Child-Centered Proactive Assistance}
A key challenge in proactive AI support for children is determining how to provide help at the right moment without requiring children to explicitly ask for it. Our findings suggest that gaze is a promising signal for this purpose, \zekun{as it provides a low-effort indication of children’s moment-to-moment visual attention.} For \textit{RQ1}, we found that during picture exploration, gaze-based proactive assistance kept children’s attention on the primary AOI for longer and guided them more frequently towards the secondary AOI than baseline assistance. Gaze metrics also indicated longer, more sustained engagement with the visual content that was talked about. For \textit{RQ2}, while there were some reservations about the quality of speech output and the narration, particularly mentioned by parents, we found a general trend that gaze-based assistance was preferred by the majority of children and accompanying adults.

These finding suggests that without requiring a child to explicitly request assistance, the gaze-based proactive assistance system can effectively use the child’s ongoing visual attention to decide \textit{when} to speak and \textit{what} part of the picture to talk about , effectively engaging the child in the storytelling. 
In this sense, gaze-based proactivity makes the assistance more grounded in the child’s own activity and better aligned with how children engage with visual materials.
Deliberate inputs such as pointing or touch could identify a region more precisely, but they require the child to decide what they want to be told about and to interrupt their exploration to communicate it. Gaze requires no separate action and lets the assistance follow an exploration already underway, although it is a weaker signal of intent than a gesture the child chose to make.
The same holds for methods that work from the picture alone, such as visual saliency. Saliency describes where attention tends to go on average, and would produce the same narration for every child viewing a given picture. What makes a region worth narrating, however, is that this particular child has attended to it, and that they have not yet explored what lies around it. Both depend on the individual exploration history, which requires observing the child rather than the image.

More importantly, our approach also guided exploration from where the child is currently looking at in a meaningful way. By starting from the child’s current focus and then introducing another relevant part of the scene, the assistant can support a smoother exploration process that is still aligned with a child's intuitive exploration. This makes the interaction feel more child-led, while still allowing the AI to take initiative. We therefore see the combination of eye tracking and LLM-based narration as a promising design direction for children’s picture book exploration, because it allows proactive assistance to remain timely, relevant, and aligned with the child’s unfolding exploration.

\subsection{Where Gaze-Based AI Assistance Fits in Practice}
Going beyond  prior work that incorporates human gaze as a behavioral signal to inform LLMs about users’ current visual focus within a task context \cite{konrad2024gazegpt,wang2024g,rekimoto2025gazellm}, 
our study shows that gaze can inform not only what an assistant talks about, but also when it speaks.
Beyond using gaze to extract a small crop of the user’s current focus, we continuously use gaze information to identify \emph{when} proactive assistance should be provided and \emph{what} the user is currently, or may soon attend to. 
Thus, gaze can function as a timely grounding signal for proactive AI assistance, enabling an implicit and low-effort interaction paradigm that is especially promising for AI systems operating in visually rich contexts.

What Ollie has achieved here could naturally extend to children’s recreational and learning activities, where proactive assistance may be especially beneficial because children often express curiosity, confusion, or interest through gaze before they can articulate these states verbally. For example, in drawing, puzzle solving, block building, museum visits, or AR-based exploration, a gaze-aware assistant could notice what a child is focusing on and provide a timely hint, question, explanation, or narrative prompt.

Beyond children’s scenarios, the same idea could extend to visually intensive activities for adults, including data analysis, document reading, design review, medical-image inspection, AR-based maintenance, and so on. In these contexts, gaze does more than indicate the visual region to pass to an LLM. It can also reflect the user's real-time states when doing the activity such as searching, comparing, hesitation, uncertainty, sustained interest, or missed information. However, these states are not equally receptive to interruption. Our method treats stable attention as a suitable moment to assist, but longer fixations have also been associated with higher cognitive load \cite{das2024, Meghanathan2015}, so a moment of stable attention may be one in which the user is actively processing what they see. Determining when a user is not only attending but also available for assistance will require signals beyond fixation stability. Used with such signals, a gaze-based proactive assistant could decide not only when a user is attending, but when they are open to being addressed,  making AI assistance more timely, situated, and minimally disruptive.

\subsection{Child Agency in Proactive Interaction}
Our findings also suggest that gaze-based proactive assistance can support child agency by making children’s ongoing attention influence how the interaction unfolds \cite{brod2026agency,borchers2026decides}. In child-centered interactions, agency is not always expressed through explicit choices; children may also communicate interest, uncertainty, or curiosity through gaze and other non-verbal behaviors. By using gaze as input, Ollie could respond to children’s emerging attentional focus rather than relying only on spoken requests or predetermined system choices. \zekun{In this sense, gaze made children’s ongoing exploration consequential for the interaction: Ollie’s support was not predetermined by the system alone, but was shaped by where children were already directing their attention. This is also reflected in participants’ feedback: several children described the gaze-based version as better aligned with their interests or giving them a stronger sense of control, and parents often perceived it as allowing the child to be more active in guiding the experience.}


At the same time, making children’s attention consequential represents only a partial form of agency-supportive design. To fully support child agency, proactive assistance should not only adapt to children’s current attentional focus, but also create opportunities for children to notice, evaluate, and redirect that support over time. This connects the present study to established theoretical views of self-regulated learning, where learners gradually develop the capacity to monitor their own state, control their actions, and adjust strategies in response to task demands \cite{schunk2003self}. In line with recent discussions of self-regulation in the context of AI in education, Molenaar’s concept of hybrid human-AI regulation offers a useful framing for this progression: AI can initially scaffold regulation by detecting learners’ needs and providing timely support, but control should gradually shift back to learners as their self-regulated learning skills develop \cite{molenaar2022concept,molenaar2022towards}. In this sense, Ollie’s gaze-based assistance can be viewed as an early form of hybrid regulation, where the system helps guide children’s exploration based on their attention, but future designs should increasingly invite children to confirm, question, revise, or redirect the system’s suggestions. 


\subsection{AI use in Children's Storytelling, Reading, and Picture book Exploration}
Our study also contributes to the broader discussion of how AI can be used in children's reading and storytelling activities. Much of prior work on AI use for children's reading and storytelling (referred to as \textit{interactive storytelling}) has focused on conversational agents that act on children's verbal input. While LLMs have been proposed as a solution to support natural, flexible, and potentially engaging conversations in interactive storytelling, it is often reported that conversational agents (with/without LLMs) for reading and storytelling exercise limited capabilities primarily due to children's limited skills of questioning, prompting, and expressing interest verbally \cite{sun2024exploring}. 

Our approach demonstrates that it is promising to use gaze to identify where a child's attention is currently directed and provide proactive guidance in children’s picture book reading activity. Specifically, gaze-based assistance successfully redirected children's attention and facilitated broader picture exploration, suggesting that AI can help structure visual engagement without requiring explicit instruction from the child. Further, the proactive support using gaze information was perceived positively by most of our participants. This finding suggests that gaze can be used to ground proactive assistance in what a child is currently attending to, without requiring them to articulate what they want.


We, however, note an important finding from our study: some participants, particularly parents, expressed hesitation about using Ollie because it would introduce yet another screen-based activity on top of children’s already-enough screen time, aligned with prior work on parents' preferences regarding screen time for their children \cite{chong2023exploring}. Further, parents emphasized the value of parent–child reading interactions and were concerned that AI support might reduce opportunities for shared engagement \cite{zhang2022storybuddy}. This insight poses a challenging tension in exploring AI use for children's reading; on one hand, AI-assisted reading and exploration has certain benefits, which were also highlighted in our study, such as enhancing interest and reducing parental caregiving load. On the other hand, as stated above, stakeholders, especially parents, have concerns about sustained use of such systems. Therefore, future studies should investigate real-world, longitudinal deployment to understand how children and parents make both strategic (weighing benefits and cost of using AI support) and spontaneous (e.g, when parents have to take care of their child while working from home) choices in using AI-based assistance.

\subsection{Limitations and Future Work}
We acknowledge several limitations of the present study. First, the final sample size for the user study was modest: 22 children completed the study, 21 contributed valid gaze data after pre-processing, which limits the generalizability of our findings across ages, reading habits, and settings. Second, the AOIs in our system were manually defined in advance; while this supported controlled analysis in the current study, it limits scalability and depends on annotators’ judgments about what counts as a meaningful region in an image. Third, as also noted by parents and the teacher, Ollie currently offers only limited interactivity. Although it can provide proactive narration, it cannot answer children’s questions, explain unfamiliar words, or sustain a more conversational exchange, which reduces its usefulness in realistic reading situations. Finally, the quality of the generated narration still needs improvement, particularly in grammar, pronunciation, repetition, delay, and voice naturalness, especially when used with languages other than English.

Targeting these limitations, our future work will focus on improving the gaze-based proactive assistant system in several concrete ways. 

First, we plan to reduce the need for manually defined AOIs by introducing automatic or semi-automatic region construction. For example, recent work combining eye tracking with state-of-the-art image segmentation suggests a promising direction in which users can collect target segmentation masks simply by looking at the AOI \cite{kirillov2023segment,wang2024gazesam}. Building on this idea, future versions of Ollie could replace hand-labeled AOIs with gaze-grounded image regions generated online, which would make the system more scalable and adaptable to new images. 

Second, recent work suggests that proactive LLM assistance can be interactive and multi-round rather than one-way \cite{liu2025proactive,deng2025proactive,wu2026excuse}. For Ollie, this points to a natural next step: extending it from a proactive narrator into a more conversational assistant that can combine gaze-based guidance with other child-centered AI supports, such as question asking \cite{zhang2022storybuddy}, drawing creation \cite{zhang2022storydrawer}, AR activities \cite{cheng2025oak}, etc..

Lastly, although the LLM generally followed the prompt and produced assistance in the expected format, the quality of its output can still be improved. A useful next step is to make the prompting more adaptive to children’s real-time responses. For example, Ollie could use gaze feedback to check whether the child actually looked at the region it just mentioned. If the child followed the narration, Ollie could continue from that region; if not, it could briefly rephrase, reconnect the ignored region to the child’s current focus, or move on without over-directing the child’s attention. Future versions could also incorporate better memory management across assistance rounds. Instead of treating each narration mostly as a local response, Ollie could remember what has already been described, what the child seemed interested in, and which regions were ignored. This would help reduce repetition, maintain coherence across multiple turns or images, and make the interaction feel more responsive to the child’s exploration.
\section{Conclusion}
Compared with explicit verbal output, children often express internal states, such as interest and curiosity, implicitly through their behavior during activities such as gaze like picture exploration. As AI systems become more capable of offering timely support, these behavioral signals can provide important cues for making assistance more relevant and less dependent on explicit requests. In this work, we advanced child-oriented proactive AI assistance by using children’s gaze behavior to generate narrative descriptions grounded in what they are currently looking at and to guide their attention toward contextually relevant areas of the picture. To examine the effectiveness of our gaze-informed proactive assistant, Ollie, we compared it with a baseline condition without gaze information in a user study. The results show that gaze-informed assistance more effectively sustained children’s attention on the narrated content and guided them toward related areas of the image. It was also preferred by both children and accompanying adults, suggesting that gaze can serve as a useful interaction signal for designing proactive AI support in open-ended children’s activities. This work shows that gaze-informed AI can make proactive assistance more responsive to children’s own ways of exploring, offering a path toward AI systems that support children without requiring them to first know what to ask.


\bibliographystyle{ACM-Reference-Format}
\bibliography{imwut}










\end{document}